%% file: main.tex
\documentclass[conference]{IEEEtran}

\input{pkgs}

\input{cmds}

\begin{document}

\title{The Role of Privacy Guarantees in Voluntary Donation of Private Health Data for Altruistic Goals}

\author{\IEEEauthorblockN{Ruizhe Wang}
	\IEEEauthorblockA{University of Waterloo\\
		ruizhe.wang@uwaterloo.ca}
	\and
	\IEEEauthorblockN{Roberta De Viti}
	\IEEEauthorblockA{MPI-SWS\\
		rdeviti@mpi-sws.org}
	\and
	\IEEEauthorblockN{Aarushi Dubey}
	\IEEEauthorblockA{University of Washington\\
		aarushid@uw.edu}
	\and
	\IEEEauthorblockN{Elissa M. Redmiles}
	\IEEEauthorblockA{Georgetown University\\
		elissa.redmiles@georgetown.edu}}

\IEEEoverridecommandlockouts
\maketitle

\begin{abstract}
	\input{abstract}
\end{abstract}

\IEEEpeerreviewmaketitle

\section{Introduction}
\label{s:intro}
\input{intro}

\section{Related Work}
\label{s:back}
\input{background}

\section{Methodology}
\label{s:methods}
\input{method}

\section{Results}
\label{s:results}
\input{results}

\section{Concluding Discussion}
\label{s:dis}
\input{discuss}

\section*{Acknowledgment}
\input{ack}


\newpage
\bibliographystyle{IEEEtran}
\balance
\bibliography{ref}
\appendix
\input{appendix}
\end{document}

%% file: pkgs.tex
\usepackage{amsmath,amsopn,amsthm,wasysym,mathtools}
\usepackage{float}
\usepackage{subcaption}
\usepackage{endnotes,microtype,xspace,graphicx,fancyvrb,multirow}
\usepackage{booktabs}
\usepackage{enumitem}
\usepackage[labelfont=bf,font=small,skip=5pt]{caption}
\usepackage[ruled,linesnumbered]{algorithm2e}
\SetKwInput{kwInit}{Init}
\SetKwComment{Comment}{\textnormal{/*} }{ \textnormal{*/}}

\usepackage{fp}
\usepackage{siunitx}

\usepackage[acronym,nohypertypes={acronym,notation}]{glossaries}

\usepackage{balance}

\usepackage{tikz}
\usetikzlibrary{calc}
\usetikzlibrary{positioning}
\usetikzlibrary{fit}
\usetikzlibrary{shapes}
\usetikzlibrary{trees}
\usetikzlibrary{arrows}
\usetikzlibrary{overlay-beamer-styles}
\usepackage{xcolor}

\usepackage{changepage}

\usepackage{balance}
\usepackage{multicol}
\usepackage{etoolbox}

\usepackage{comment}
\PassOptionsToPackage{hyphens}{url}\usepackage{hyperref}
\usepackage[T1]{fontenc}

%% file: cmds.tex
\fvset{fontsize=\scriptsize,xleftmargin=8pt,numbers=left,numbersep=5pt}

\def\Snospace~{\S{}}

\usepackage{pifont}

\newif\ifdraft\drafttrue
\newif\ifnotes\notestrue
\ifdraft\else\notesfalse\fi

\input{glyphtounicode}
\newcolumntype{R}[1]{>{\raggedleft\let\newline\\\arraybackslash\hspace{0pt}}p{#1}}

\newcommand{\squishlist}{
\begin{itemize}[noitemsep,nolistsep]
  \setlength{\itemsep}{-0pt}
}
\newcommand{\squishend}{
  \end{itemize}
}

\usepackage{tikz}

\usepackage{xstring}
\newcommand{\PP}[1]{
\vspace{2px}
\noindent{\bf \IfEndWith{#1}{.}{#1}{#1.}}
}

\newcommand{\boxbeg}{
\vspace{2px}
\noindent\begin{tabular}{|l|}\hline
\begin{minipage}{3.2in}
\vspace{2px}
\noindent
}

\newcommand{\boxend}{
\vspace{2px}
\end{minipage}\\ \hline
\end{tabular}
\vspace{-10pt}
}

\def\BibTeX{{\rm B\kern-.05em{\sc i\kern-.025em b}\kern-.08em
    T\kern-.1667em\lower.7ex\hbox{E}\kern-.125emX}}
\definecolor{black}{rgb}{0,0,0}
\definecolor{Orange}{rgb}{1,0.5,0}
\definecolor{Red}{rgb}{1,0,0}
\definecolor{Green}{rgb}{0,0.8,0.5}
\definecolor{Purple}{rgb}{0.75,0,1}
\definecolor{babypink}{rgb}{0.96, 0.76, 0.76}
\definecolor{cerulean}{rgb}{0.0, 0.48, 0.65}
\definecolor{azure}{rgb}{0,0.49,1}
\definecolor{periwinkle}{rgb}{0.8, 0.8, 1.0}
\definecolor{Pink}{RGB}{255, 102, 204}
\definecolor{electriccyan}{rgb}{0.0, 1.0, 1.0}
\definecolor{dodgerblue}{rgb}{0.12, 0.56, 1.0}

\newif\iflongversion
\longversiontrue   

\newcommand{\twovers}[2]{\iflongversion #2\else #1\fi}

%% file: abstract.tex
While voluntary donation of private health information enables valuable research, privacy concerns often deter potential donors. Privacy Enhancing Technologies (PETs) aim to address these concerns, yet their effectiveness in encouraging data sharing remains unclear. 
This study conducts a vignette survey ($N=494$) with participants recruited from Prolific to examine the willingness of US-based people to donate medical data for developing new treatments. It investigates four general guarantees offered across PETs: data expiration, anonymization, purpose restriction, and access control and two mechanisms for verifying these guarantees: self-auditing and expert auditing. This study also controls for the impact of confounds, including demographics and two types of data collectors: for-profit and non-profit institutions.

Our findings reveal that 
respondents hold such high expectations of privacy from non-profit entities a priori that explicitly outlining privacy protections has little impact on their overall perceptions.
In contrast, offering privacy guarantees elevates respondents' expectations of privacy for for-profit entities, bringing them nearly in line with those for non-profit organizations. 
Further, while the technical community has suggested audits as a mechanism to increase trust in PET guarantees, we observe limited effect from transparency about such audits. We emphasize the risks associated with these findings and underscore the critical need for future interdisciplinary research efforts to bridge the gap between the technical community's and end-users' perceptions regarding the effectiveness of auditing PETs. 

%% file: intro.tex
The altruistic use of personal health data holds immense potential for societal benefit. For example, the public health sector can leverage individuals' medical histories and health data to analyze epidemiological trends, enhance disease surveillance, improve risk prediction models, diagnose rare and emerging diseases, and accelerate the development of novel treatments~\cite{Bietz2015, Aitken2016, Bartlett2018, GREENE201949, Kim2015, Trinidad2020}.

Beyond direct medical applications, health data sharing offers economic benefits. A unified data-sharing infrastructure could eliminate redundant clinical trials and research efforts, thereby reducing healthcare costs~\cite{szarfman2022recommendations}. According to the European Commission, innovations that enhance health data accessibility could generate savings of billions of euros within the European healthcare sector alone~\cite{digital-strategy}.
These demands are further intensified by recent advances in artificial intelligence, particularly machine learning models, whose effectiveness heavily depends on the quality and quantity of data they are trained on.

Recognizing these converging medical, scientific, and economic imperatives, the National Academy of Medicine (NAM) has advocated for more open and collaborative approaches to health data sharing~\cite{whicher2020health}, emphasizing both the ethical obligation and practical necessity of broader data access frameworks.

However, despite these compelling benefits and institutional support for expanded health data sharing, privacy concerns remain a significant barrier to widespread data donation.
Personal health data is routinely collected by healthcare providers~\cite{Bartlett2018, Kim2015}, mobile applications, and wearable devices~\cite{GREENE201949}, often flowing to third parties through data-sharing agreements~\cite{Trinidad2020} with limited transparency to the individuals who generated it.
Studies consistently show that privacy concerns are a primary deterrent to individuals contributing their sensitive health information~\cite{DGOV20, voigt2020willingness, silber2022preregistered, GREENE201949, Kim2015}, particularly when hospitals permit data sharing with third parties without explicit patient consent~\cite{leave-hospital}.

In an effort to foster data collection and analysis while protecting the privacy of data contributors, there has been an increasing focus on deploying \textit{privacy-enhancing} technologies (PETs) for data storage~\cite{popa2011cryptdb,blindseer}, data processing~\cite{Arasu2013, Bajaj2011, Baumann2015,Gupta2019, Chowdhury20,al2019privacy,anonify}, and machine learning (ML) training~\cite{wuhybrid, de2022covault}.
Some systems explicitly consider data donation scenarios: Waldo~\cite{dauterman2022waldo} enables privacy-preserving queries over medical time-series data for remote patient monitoring; CoVault~\cite{de2022covault} supports expressive queries on sensitive data at scale under a strong threat model, and considers a national-scale epidemic analytics scenario; Anonify~\cite{anonify} provides decentralized dual-level anonymity for medical data donation; and Mycelium~\cite{roth2021mycelium} supports differentially-private queries over large distributed graphs (e.g., disease-spread data) across millions of user devices.

However, despite these technological advances, implementing PETs alone does not guarantee increased data availability. Privacy protections must be expressed to potential donors in comprehensible terms that address their fundamental concerns, or these sophisticated mechanisms remain ineffective at encouraging data sharing.

In order to evaluate whether the guarantees offered by PETs ultimately impact people's willingness to donate their data, at least three steps are necessary. First, we must be able to effectively explain guarantees (either general guarantees or those of a particular PET) to the people they aim to protect: end-users. Prior work has focused on doing so for particular PETs, mainly differential privacy (DP) and end-to-end encryption (E2EE)~\cite{CCS21a,abu2018exploring,WPES21,bai2019user,de2016expert,dechand2019encryption,abu2017obstacles,wu2018tree,xiong2020towards,nanayakkara2023chances}. Second, using these explanations, we must evaluate whether the guarantees impact people's privacy concerns (or expectations), and in turn, their willingness to share data. Prior work has done such evaluations using self-report studies focused on individual PETs, again primarily DP~\cite{CCS21a, bullek2017towards,nanayakkara2023chances} or E2EE~\cite{dechand2019encryption, bai2020improving, abu2018exploring}. Third, while rare, we must validate those results in the field to confirm that the effects observed replicate when studying actual versus intended behavior (e.g., \cite{chi22, acquisti2013privacy}).

In this work, we focus on addressing the first two steps outlined above: effectively explaining privacy guarantees and evaluating their impact on data-sharing intentions. We follow real-world medical consent frameworks~\cite{practice3,practice2} and privacy regulations~\cite{neame2013effective,kazley2014association}, and individually investigate four fundamental guarantees across different technologies: anonymization, access control, data expiration, and purpose restriction (PG(1)--PG(4)), along with two verification processes: expert and self auditing (AG(1)--AG(2)). This approach helps identify which specific protections most influence privacy expectations and data-sharing decisions.

We examine the role of these PET guarantees and their audits in shaping privacy expectations and influencing data-sharing intentions within a specific context used in prior work (see e.g., \cite{CCS21a,nanayakkara2023chances}): medical data donation.
We ask: 
\begin{enumerate}[
	leftmargin=*,
	label={RQ(\arabic*):},
	ref={RQ(\arabic*)}]
	\item How well do people understand and expect what is offered by the privacy-preserving guarantees PG(1)--PG(4) and auditing guarantees AG(1)--AG(2)?
	\item How does the deployment of privacy guarantees and auditing influence people's willingness to donate their personal health data?
\end{enumerate} 

To address these questions, we conduct a vignette survey ($n=494$) following the best practice methodology in prior work~\cite{CCS21a,abu2018exploring,xiong2020towards}. Each survey respondent is presented with a hypothetical opportunity to donate their health data to help develop a treatment for a specific chronic disease, along with how their data will be protected (PG(1)--PG(4) enforced (or not) by AG(1)--AG(2)). 
We control for confounding factors identified in prior work, including data-collection entity~\cite{Trinidad2020, HendricksSturrup2023} and socio-demographics~\cite{hargittai2020americans,dooley2022field, Trinidad2020}.

We find that even when told nothing about PETs implemented by the entity, participants are 23\% more likely, on average, to expect a non-profit to implement PG(1)--PG(4) and AG(1)--AG(2). 
As a result of these already high privacy expectations for non-profit organizations, we find that mentioning a specific privacy protection in the survey does not significantly enhance people's willingness to donate towards non-profit entities: even when no privacy protection is explicitly mentioned, 89\% of the participants are willing to donate to a non-profit entity. 
In contrast, for-profit entities need to effectively \textit{demonstrate} their privacy protections; indeed, explicitly mentioning privacy protections in the survey does increase privacy expectations of for-profit entities from 50\% to the level of non-profit entities. Privacy expectations, in turn, influence the willingness to donate. 

Furthermore, while the technical community has suggested \textit{external audits} as a mechanism to increase trust in PET implementation, our initial inquiry suggests that more work is needed to explain the purpose and effectiveness of such audits to end-users. 
In fact, the effect of audit statements on people's willingness to donate is limited to a specific scenario involving for-profit entities and auditing to check that purpose restriction (PG(4)) is correctly implemented.

We argue that it is critical for non-profit entities to rigorously implement data privacy measures , as any future data leak could lead to a significant loss of trust. Prior research~\cite{acquisti2013privacy} underscores that users place greater value on \textit{maintaining} their expected privacy than on \textit{gaining} additional privacy they did not initially anticipate. In contrast, we highlight the risk of for-profit entities engaging in ``privacy washing''~\cite{cirucci2024oversharing}, where statements about PETs are used to artificially raise privacy expectations to encourage data collection. At the same time, our findings reveal that respondents are perceptive of the limitations in general PET guarantees, particularly concerning protections against data breaches. This underscores the need for future research to explore how to effectively communicate the potential of stronger, emerging PET guarantees and address the skepticism of end-users.

\twovers{The \href{https://arxiv.org/abs/2407.03451}{full appendix} of the paper is available.}{}

%% file: background.tex
\label{sec:background}
In this section, we examine prior work on people's willingness to donate personal data, privacy concerns, the role of PETs, and educational or explanatory strategies for PETs.

\medskip
\PP{Data donation} The analysis of personal health data is crucial for medical research, particularly in diagnosing emerging or rare diseases and developing new treatments. Indeed, the COVID-19 pandemic has further intensified the demand for personal health data~\cite{scott2020advanced}. Despite the importance of this data, the sensitive nature of health data poses significant obstacles to its collection~\cite{fahey2020covid}. Prior work indicates that individuals are generally willing to donate their data for altruistic purposes~\cite{pilgrim2022, Aitken2016, seberger2021post}, though there is a notable reluctance when it comes to their health data specifically~\cite{Middleton2020, mori2016public}. This reluctance diminishes when the donor or their close family members are directly affected by the disease under study~\cite{Goodman2017, Bartlett2018}, suggesting that non-privacy-related factors influence donation decisions.

\medskip
\PP{Privacy concerns} Concerns about privacy and the misuse of donated data are prominent among potential donors~\cite{mori2016public}. These concerns include the risk of being identified, discriminated against, and having personal sensitive data misused or leaked~\cite{Weitzman2012, Aitken2016, Kim2015, McCormick2021}. A common source of these concerns is distrust of the receiving entity~\cite{Weitzman2012, Howe2018, Aitken2016, Stockdale2019}, often due to fears that the entity might share data with unauthorized third parties~\cite{Kacsmar2022, Middleton2020}. This distrust is intensified when participants are unfamiliar with the recipient entity~\cite{Weitzman2012, Howe2018}, or if it is a governmental or for-profit organization~\cite{Aitken2016, Stockdale2019, Garrison2016, HendricksSturrup2023, khanra2023bridging, amasurvey}.
Beyond this \textit{intentional} data misuse, there are concerns about data leakage caused by hackers or unintentional mishandling~\cite{seh2020}. People are reluctant to interact with entities having any history of data breaches, doubting their ability to safeguard sensitive information~\cite{seh2020}.

\medskip
\PP{Mental models on PETs}
PETs aim to address user privacy concerns by providing technical safeguards for sensitive data. While various PETs are available for health data donation contexts~\cite{jordan2022selecting}, their effectiveness depends largely on users' trust and understanding. Prior work indicates that participants with greater online privacy literacy tend to have more trusting attitudes in PETs~\cite{harborth2020privacy}, while those with limited knowledge often remain skeptical.

Studies have also found that non-experts systematically misunderstand privacy technologies~\cite{wu2018tree, Oates2018, dechand2019encryption}. For instance, many incorrectly conceptualize encryption as merely a form of access control~\cite{wu2018tree} or confuse it with simple data encoding~\cite{abu2017obstacles}. Additionally, non-expert end users may struggle to understand the consequences of inadequate privacy protections~\cite{tang2021defining, lev2022, abu2018exploring, gerber2019, pham2019moral, vaniea2014mental}. 
Lerner et al.~\cite{Lerner2022} identified an even more fundamental barrier: some participants express inherent skepticism about the existence of ``true'' privacy, illustrating how deep-seated misconceptions can fundamentally undermine the reassurance PETs are designed to provide.

As a result, even when privacy protections are present, people may have risk expectations that are misaligned with reality. We investigate the alignment between stated privacy guarantees and people's expectations for how their data will be protected as part of RQ1.

\medskip
\PP{Explaining PETs}
To address these misunderstandings and evaluate the impact of PETs on downstream factors such as privacy concerns (or expectations) and willingness to share data, technologists and researchers have worked to explain PETs to the public to address their unfamiliarity with privacy concepts~\cite{cranor2006}.
Prior work focuses heavily on E2EE and DP~\cite{abu2018exploring, bai2019user, CCS21a, de2016expert,WPES21,dechand2019encryption,abu2017obstacles,wu2018tree,xiong2020towards,nanayakkara2023chances}. These efforts remain ongoing, as effectively and scalable setting privacy expectations remains a challenge. Methods found effective to explain PETs include visualizations~\cite{stransky2021limited}, mental models~\cite{stewart2012death, lin2012expectation}, nutrition labels~\cite{kelley2010standardizing}, metaphors~\cite{karegar2022exploring}, short statements~\cite{papacharissi2005online}, and privacy games~\cite{privgame2018, edgame2023}. 
However, there is a gap in the literature regarding techniques to explain auditing guarantees, despite the importance of auditing in verifying compliance with privacy promises.

\medskip
\PP{Privacy guarantees (PGs) and auditing}
Prior research on PETs and their accompanying PGs has primarily focused on evaluating specific technologies (e.g., DP~\cite{valdez2019users,CCS21a} or E2EE~\cite{dechand2019encryption}) and their combined impact on privacy concerns. A smaller subset of studies has assessed how these technologies affect willingness to share data~\cite{valdez2019users}. Additional research has examined public understanding of specific guarantees such as data retention~\cite{deletionusers}, data anonymization~\cite{haddow2011nothing}, and secondary use permissions~\cite{juga2021willingness}.

Our approach differs by systematically comparing four core PGs implemented across many PETs: data expiration, data anonymization, use restriction, and access control (see~\autoref{s:intro}). We also separately investigate auditing mechanisms, which serve as verification procedures rather than direct guarantees.
In particular, we examine whether offering a given PG influences people's willingness to donate health data to the recipient entity. Furthermore, we focus on the effect of two different auditing processes -- expert auditing and self auditing -- on enhancing privacy expectations and willingness to share data. 

To our knowledge, this is the first study to systematically compare these four core PGs and evaluate how different auditing mechanisms affect user privacy expectations and willingness to donate health data.

%% file: method.tex


As mentioned in~\autoref{s:intro}, we address research questions RQ1 and RQ2 through a user survey.
In this section, we discuss the ethical considerations of our study (\autoref{ss:method-ethics}), outline the explored statements (\autoref{ss:method-scenarios}) and survey design (\autoref{ss:method-design}), and describe the cognitive interview and pilot study process used to refine and improve participants' comprehension of the presented scenarios (\autoref{ss:method-understand}).
We also describe our participant selection strategy (\autoref{ss:method-recruit}), analysis procedure (\autoref{ss:analysis}), and the limitations of our methodology (\autoref{ss:method-limit}).


\subsection{Ethics}
\label{ss:method-ethics}

Our study was conducted under approval from our university's Ethics Review Board (ERB). The approval covered both the cognitive interviews and the survey. We implemented several measures to ensure ethical treatment of participants:

All survey data was collected anonymously through Qualtrics~\cite{qualtrics}, with IP address collection disabled.
Participants were presented with a consent form at the beginning of the survey that detailed the study purpose, data handling procedures, and compensation details. Participants could opt out of answering any demographic questions without affecting their compensation.

For cognitive interviews conducted via Zoom~\cite{zoom}, participants were asked to avoid using their real names when joining the meetings to maintain anonymity. Only anonymized transcripts were retained for analysis after removing any personal identifiable information.

All research participants were recruited anonymously through Prolific~\cite{prolific}, providing an additional layer of separation between researchers and participants' identities. The collected data was accessible only to the research team, and all data will be deleted after the completion of the research project in accordance with our ERB protocol.

Each participant received $1.20\$$ upon completing the survey.
To determine this figure, we ran a test of 20 participants, who took 5m56s (median) to complete it. Thus, the actual compensation was $12\$$/hr, which aligns with Prolific recommendations. 
Interview participants were first recruited through a screening survey that collected demographic information and assessed willingness to participate in interviews. 
Initially, we compensated participants 1.25$\$$ for completing this 5-minute screener (approximately 13.68$\$$/hr), titled ``Sign Up for Paid Interview on Data Donation.'' However, this above-average compensation led to only 8\% interview attendance, as many users were motivated primarily by the screening fee.

To better align incentives and improve participation rates, we adjusted the screener compensation to 0.83$\$$ (approximately 9.96$\$$/hr), which better reflected platform norms for screening surveys. This adjustment increased our interview attendance rate to 21\%. Participants who completed the 30-minute interviews were additionally compensated 15$\$$.

\subsection{Donation Scenarios}
\label{ss:method-scenarios}

\begin{figure*}[h]
        \centering
		\begin{tabular}{|p{.95\textwidth}|}
			\hline
			\begin{tabular}{@{}p{.95\textwidth}@{}}
			\textbf{PG(1) Anonymization}: The privacy-preserving technology removes any personal identifiable information at the time of data collection, so that the data stored by the recipient entity is not linkable to its data owner (i.e., the data is anonymous).
			\end{tabular} \\
			\hline\hline
			\begin{tabular}{@{}p{0.95\textwidth}@{}}
				
			\textbf{PG(2) Access control}: The privacy-preserving technology restricts data access to the authorized scientists (within the recipient entity) which are working on the treatment for the given disease.
			\end{tabular} \\
			\hline\hline
			\begin{tabular}{@{}p{0.95\textwidth}@{}}
			\textbf{PG(3) Data expiration}: The privacy-preserving technology discards the donated data, or makes it inaccessible after a given expiration time, which can be chosen by the data donor in the data collection agreement.
			\end{tabular} \\
			\hline\hline
			\begin{tabular}{@{}p{0.95\textwidth}@{}}
			\textbf{PG(4) Purpose restriction}: The privacy-preserving technology ensures that the recipient entity can only use the donated data to develop a treatment for the given disease, and not in the context of any other research they may be working on (e.g., different disease).
			\end{tabular} \\
			\hline\hline
			\begin{tabular}{@{}p{0.95\textwidth}@{}}
			\textbf{Baseline: N/A}: (No privacy guarantee is at all mentioned.) 
			\end{tabular} \\
			\hline
		\end{tabular}
		\caption{Privacy statements PG(1)--PG(4). One of the five statements (including the empty baseline statement) is randomly presented in the donation scenario immediately after the survey introduction.}
		\label{fig:privacystatements}
	\end{figure*}
	
	\begin{figure*}[h]
            \centering
			\begin{tabular}{|p{.95\textwidth}|}
				\hline
				\begin{tabular}{@{}p{.95\textwidth}@{}}
				\textbf{AG(1): Expert auditing}: An external advisory board of scientists and software engineers appointed by the recipient entity will regularly verify that the privacy-preserving technology is working as described. The results of this verification will be made public.
				\end{tabular} \\
				\hline\hline
				\begin{tabular}{@{}p{0.95\textwidth}@{}}	
				\textbf{AG(2): Self-auditing}: Anyone interested, including the respondent and the experts the respondent trusts, will be able to verify that the privacy-preserving technology is working as described. Anyone, including the respondent and the experts the respondent trusts, can make their verification results public.
				\end{tabular} \\
				\hline\hline
				\begin{tabular}{@{}p{0.95\textwidth}@{}}
				\textbf{Baseline: N/A}: (No auditing statement is at all mentioned.)
				\end{tabular} \\
				\hline
			\end{tabular}
			\caption{Auditing statements AG(1)--AG(2). One of the three statements (including the empty baseline statement) is randomly presented in the donation scenario after a non-control \texttt{privacy statement}.
			}
			\label{fig:auditingstatements}
		\end{figure*}
Many PETs involve complex mechanisms that cannot be adequately explained in brief consent forms without compromising informed consent principles. Additionally, consent forms must satisfy legal and ethical requirements while remaining accessible, as participants typically spend just minutes reviewing them~\cite{de2022using,nathe2019challenges,collectingsurvey}.

To address this challenge, we align with real-world consent practices~\cite{practice2,practice3} by measuring reactions to the high-level guarantees that PETs provide rather than explaining their technical implementations. This approach allows us to investigate fundamental privacy guarantees that appear consistently across different technologies—guarantees that are both meaningful to users and reflective of actual implementations:
\begin{enumerate}[
	leftmargin=*,
	label={PG(\arabic*):},
	ref={PG(\arabic*)}]
	\item Anonymization \cite{8819477, shahid2022two}: data is not linkable to its owner (as defined in \cite{chevrier2019use});
	\item Access control \cite{GREENE201949, hu2006assessment, ni2010privacy, tourani2017security}: data is accessible only by authorized people, specified in the data collection agreement;
	\item Data expiration \cite{10.1093/cybsec/tyy001, 8726847, 1628485}: data is discarded or inaccessible after  a given expiration time, specified in the data collection agreement;
	\item Purpose restriction \cite{de2022covault, byun2005purpose}: data is only used for the stated collection purpose, specified in the data collection agreement;
\end{enumerate}

In practice, implementing all guarantees simultaneously can be costly and challenging for real-world systems~\cite{neame2013effective,kazley2014association}.
This constraint is reflected in many consent frameworks, which often focus on providing a single guarantee—such as anonymization~\cite{practice2} or access control~\cite{practice3}.
Following this practical approach, our study investigates the impact of each guarantee \textit{individually}, rather than presenting them in combination.
This targeted analysis enables us to identify \textit{which} specific protections most significantly influence privacy expectations and, consequently, reported data-sharing intentions.

Beyond these privacy guarantees, the technical community has developed methods to verify that PETs function as intended. We also investigate how these auditing mechanisms affect user trust and willingness to donate data, specifically examining two approaches\footnote{We considered but excluded government auditing as a separate category due to overlap with ``expert auditing'' and potential political bias (see~\autoref{sec:background}).}:
\begin{enumerate}[
        noitemsep,leftmargin=\parindent,
	leftmargin=*,
	label={AG(\arabic*):},
	ref={AG(\arabic*)}]
	\item Expert auditing~\cite{li2015secure,wang2011privacy}: engaging an aggregator-selected external advisory board to audit the system to verify that the PET functions as described in the data collection agreement;
	\item Self auditing~\cite{holt2005logcrypt,chen2022blockchain}: granting anyone, including donors or external advisors appointed by them, the ability to perform such audits.
\end{enumerate}

We present all statements in plain text format, avoiding metaphors and visual aids due to their mixed effectiveness in privacy communications~\cite{peekinginto,Demjaha2018} and to align with standard practice in medical and survey consent forms~\cite{practice3,practice2}. 
\autoref{fig:privacystatements} and \autoref{fig:auditingstatements} show the exact text presented (e.g., the factor levels).

Justification on selecting the included PETs is in~\autoref{app:justification}.

\subsection{Survey Structure}
\label{ss:method-design}

\begin{figure}[ht]
	\centering
	\includegraphics[width=.47\textwidth]{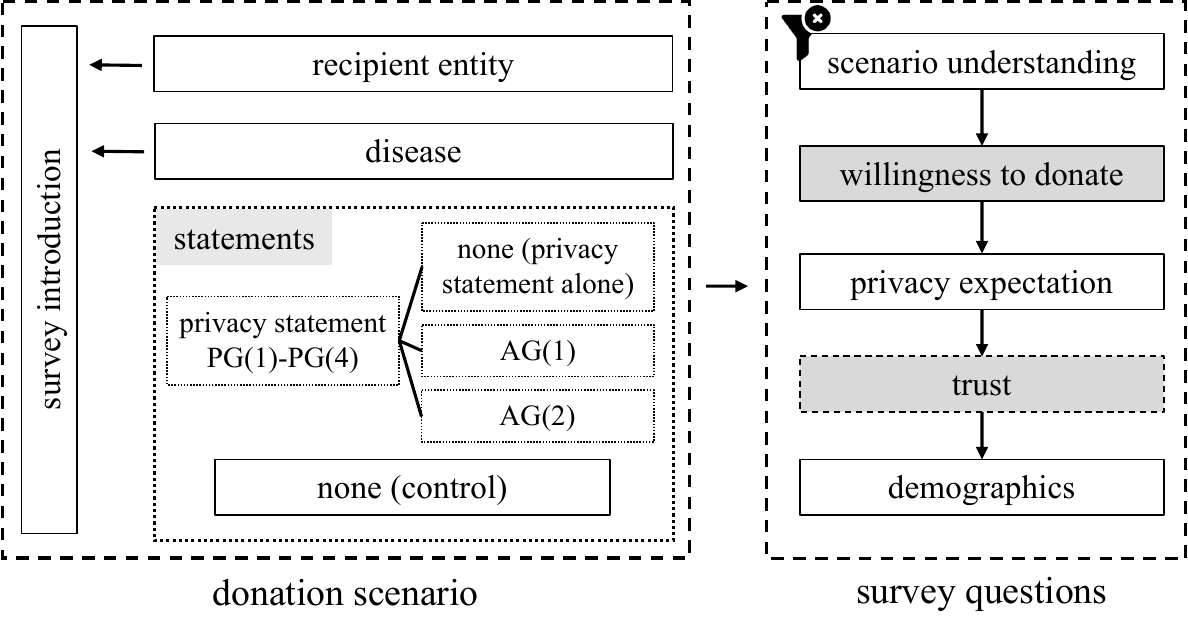}
	\caption{Survey design. Each participant is presented with a donation scenario and a survey introduction with two variations: recipient entity and disease. \texttt{statements} (PG(1)--PG(4) and AG(1)--AG(2)) are also presented in the donation scenario. Participants are then asked five sets of questions (survey questions). When the control statements are presented, the \texttt{trust} (dashed) is omitted. The \texttt{willingness to donate} and \texttt{trust} questions (dark-shaded) are associated with an open-text question. The \texttt{scenario understanding} is used as a filter. Participants who did not understand the donation scenario were excluded from the analysis. A sample screen that participants would encounter is provided in \autoref{fig:sample}.}
	\label{fig:flow}
\end{figure}

Our survey design is shown in~\autoref{fig:flow}, and the full survey can be found in~\autoref{app:survey}. 
First, we present to the respondents a \textit{suvery introduction}, which we detail in~\autoref{fig:scenario}. 
\begin{figure*}[ht]
        \centering
		\begin{tabular}{|p{.95\textwidth}|}
			\hline
			\textbf{Survey introduction} \\
			\begin{tabular}{@{}p{.95\textwidth}@{}}
				Imagine that an \textit{entity} wants to develop a new treatment
				for \textit{disease}. They need medical data from people with and without
				\textit{disease} to develop the treatment. They ask you to donate your medical record to
				help develop the treatment.
				Your medical record contains your: (i) personal information, which may include information
				about your age, weight, gender, race; (ii) medical history, which may include information about
				allergies, illnesses, surgeries, immunizations, and results of physical exams and tests; and (iii)
				medical behavior, which may include information about medicines taken and health habits, such
				as smoking habits, diet and exercise.
			\end{tabular} \\
			\hline
		\end{tabular}
		\caption{Survey Introduction. The \textit{entity} type (for-profit or non-profit) and the \textit{disease} (selected from a list of common chronic diseases) are randomly selected.}
		\label{fig:scenario}
	\end{figure*}
We present a \textit{donation scenario} using this introduction: we want to assess respondents' willingness to donate their health data to a \emph{recipient entity} developing a new treatment for a specific \emph{disease}. The type of entity (for-profit or non-profit) and the disease (cancer, diabetes, heart failure, high blood pressure, and stroke) are randomly selected; the disease options are taken from a list of common chronic diseases published by the Centers for Medicare and Medicaid Services (CMS)~\cite{cms-chronical}.

Then, we present to the respondents either a control (no statement) or an experimental statement. Experimental statements are composed of either one privacy statement, alone, or one privacy statement and one auditing statement. Privacy statements are uniformly selected at random from a pool of four privacy statements; auditing statements are selected at random from two auditing statements. Specifically, one statement from~\autoref{fig:privacystatements} and one from~\autoref{fig:auditingstatements} are presented to each respondent.

%

%

Our {survey questions} assess respondents' self-reported:
\begin{itemize}
 \item \texttt{scenario understanding}, used to filter out respondents that report not understanding the scenario (see~\autoref{ss:method-recruit}): ``How would you rate your understanding of the above scenario?'' (4 point Likert scale: Fully understand - Not understand).
 \item \texttt{willingness to donate} their health data to the recipient entity: ``In this scenario how likely would you be to donate your medical record?'' (4 point Likert scale: Very likely - Very unlikely).
  \item \texttt{privacy expectations} regarding the specific privacy guarantees we investigate, measured by their agreement with the statements presented in~\autoref{fig:sqconcerns} using the same Likert scale as \texttt{willingness to donate}.\footnote{This question included an additional attention check with a sixth statement: ``If I donate my data, I will meet Albert Einstein." Respondents who did not answer `Very Unlikely' were removed from the experiment, as detailed in~\autoref{ss:method-recruit}.} We assess participants' expectations about all guarantees, regardless of what statement they were presented. 
 \item \texttt{trust} that the recipient entity will implement the protection described by the \texttt{privacy statement}: ``I trust the \textit{entity} will handle my data as described.'' (4-point Likert scale Strongly agree - Strongly disagree).
\item demographics and experiences, including age, gender, education level, donation history,
technical background (i.e., education background and job field relationships with computer technologies), and egocentricity(i.e., whether respondents or their close relatives have the disease, as defined in~\cite{grgic2020dimensions, USENIXSec17}).


\end{itemize}

\begin{figure*}[h]
        \centering
		\begin{tabular}{|p{.95\textwidth}|}
			\hline
			\textbf{Anonymization}\\
			\begin{tabular}{@{}p{.95\textwidth}@{}}
				My full name or other personal identifiable information will be linked to the donated medical record.
			\end{tabular} \\
			\hline\hline
			\textbf{Access control}\\
			\begin{tabular}{@{}p{0.95\textwidth}@{}}
				Any employee at the recipient entity will be able to access the donated medical records.
			\end{tabular} \\
			\hline\hline
			\textbf{Data expiration}\\
			\begin{tabular}{@{}p{0.95\textwidth}@{}}
				The donated medical record will be deleted at a set point in time.
			\end{tabular} \\
			\hline\hline
			\textbf{Purpose restriction}\\
			\begin{tabular}{@{}p{0.95\textwidth}@{}}
				The donated medical records will be used for another purpose without my consent.
			\end{tabular} \\
			\hline\hline			
			\textbf{Expert auditing}\\
			\begin{tabular}{@{}p{0.95\textwidth}@{}}
				A group of independent experts will verify whether the privacy-preserving technology works and publish a report on their findings.
			\end{tabular} \\
			\hline\hline
			\textbf{Self-auditing} \\
			\begin{tabular}{@{}p{0.95\textwidth}@{}}
				I will be able to hire someone to verify that my medical record is protected as described.
			\end{tabular} \\
			\hline
		\end{tabular}
		\caption{To measure \texttt{privacy expectations} respondents reported their agreement with each statement listed above on a 4-point Likert Scale from ``Very Likely'' to ``Very Unlikely''.
		}
		\label{fig:sqconcerns}
	\end{figure*}

\begin{figure}[]
	\begin{subfigure}{.47\textwidth}
		\includegraphics[width=\textwidth]{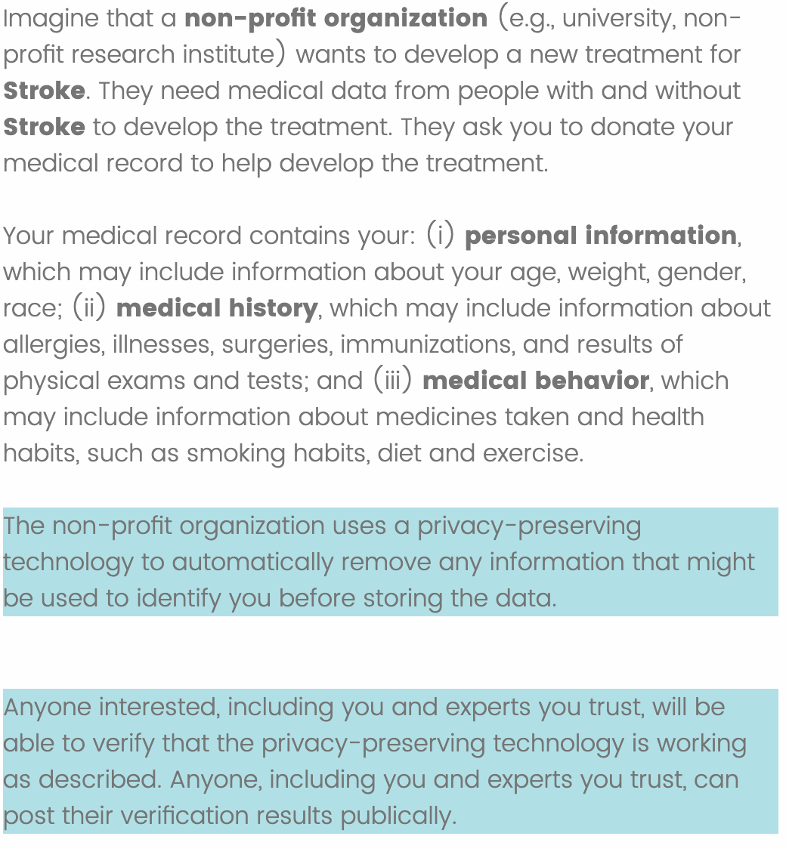}
		\end{subfigure}
		\begin{subfigure}{.47\textwidth}
			\includegraphics[width=\textwidth]{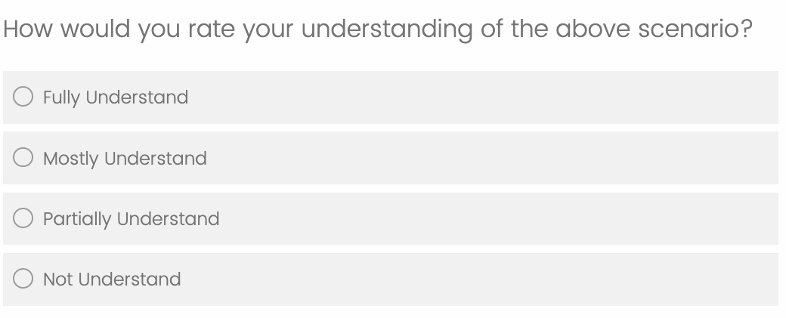}
			\end{subfigure}
\caption{Screenshot of the survey, which consists of three segments. The first segment presents the \textit{survey introduction}, varying the \textit{disease} and the \textit{entity} across participants. The second segment presents the randomized \textit{privacy statement} and \textit{auditing statement}, highlighted with color coding for emphasis. The third segment poses the survey question that participants are required to answer.}
\label{fig:sample}
\end{figure}

We also ask respondents to explain their responses to the \texttt{willingness to donate} and \texttt{trust} questions. 
These explanations are collected through open-ended questions: ``Why you are willing (or unwilling) to share your medical record with this \textit{entity}?'' and ``Please explain why you do (or do not) trust that the \textit{entity} will handle your data as described.'' 
We note that the \texttt{scenario understanding} and \texttt{trust} questions are only asked when a non-control privacy statement is presented.

A screenshot of the survey is presented in~\autoref{fig:sample} to illustrate the survey design.



\subsection{Survey Refinement}
\label{ss:method-understand}
We refine our survey thourgh a pilot study and multiple iterations of cognitive interviews, following the design of~\cite{de2022using}. Inividuals are only allowed to participate in one of the studies.

We use the pilot study to test the survey design and the donation scenario. We found, for example, that no significant relationship between different diseases and donation willingness($p=0.32$, $\chi^2= 17.00$) in the pilot survey ($N=81$ in the control condition). We thus present five common chronic diseases in the final survey (see~\autoref{app:survey}) to ensure participants can relate to the donation scenario and catch positive egocentricity measurement. We list the major takeaways in~\autoref{app:pilot}.

We conducted 17 cognitive interviews with participants recruited through Prolific~\cite{prolific}, selected from 49 users who completed our screening survey and indicated their willingness to participate in the interview. These interviews evaluated participants' understanding of the survey content and helped facilitate consistent comprehension across respondents.
Each interview involved presenting the survey to participants and actively seeking feedback on their interpretation of the data donation scenario, and the privacy and auditing statements. We additionally presented all statements to the interviewees and asked for their understanding and perception of each of them. 

We continued refining the survey through cognitive interviews until no further constructive feedback was received. In the final interview, we observed generally good understanding of the statements and the donation scenario. For example, the participant described anonymization (PG(1)) as \textit{``eliminating any sort of demographic or personal information affiliated with your data that could be associated with you if the data were to somehow be leaked''} and access control (PG(2)) as \textit{``some sort of database that only certain individuals have access to by using a passcode and certain credentials to to log in.''}
%


\subsection{Survey Sampling}
\label{ss:method-recruit}
Following the recommendation of~\cite{tang2022replication}, we recruited participants through Prolific~\cite{prolific} and collected 560 responses.
To ensure focused results, we restricted respondents to adults residing in the U.S. and requested a gender-balanced distribution. Our recruitment included 272 men, 275 women, 9 nonbinary individuals, and 4 participants who chose not to disclose.

Of these 560 respondents, we sequentially excluded the following from the analysis: 
35 that submitted incomplete responses; 20 that failed our attention check~(\autoref{ss:method-design}); and 11 that indicated they did not understand the survey scenario (using \texttt{scenario understand}). 
Thus, our final dataset comprises 494 respondents.
To enhance statistical power, we limit the levels of age and education to binary groups based on the data distribution among our participants. 
We report their demographics in~\autoref{tab:demographics} and the number of respondents assigned to each condition in~\autoref{tab:scenarios}.

\begin{table}
	\centering

	\caption{Participant demographics. We note that the second category of each demographical attribute (e.g., no technical background) is considered the baseline scenario during analysis, except for ``age'', which is treated numerically.
    }
	\label{tab:demographics}
	\begin{tabular}{lcccc}
		\toprule
		Description & Category & $n$ & \% \\
		\midrule
		\multirow{4}{*}{Age} & 18 - 29 & 120 & 24.3\% \\
            & 30 - 49 & 222 & 44.9\% \\
            & 50 - 64 & 111 & 22.5\% \\
            & 65+ & 41 & 8.3\% \\\midrule
             
            \multirow{3}{*}{Gender} & Woman & 254 & 51.42\% \\
            & Man & 231 & 46.76\% \\
            & Others & 9 & 1.82\% \\\midrule
            
            \multirow{2}{*}{Education} & B.S. or above & 358 & 72.47\% \\
            & Up to H.S. & 136 & 27.53\% \\\midrule
            
            \multirow{2}{*}{Technical Background} & Yes & 129 & 26.11\% \\
            & No & 365 & 73.89\% \\\midrule
            
            \multirow{2}{*}{Donation History} & Yes & 56 & 11.34\% \\
            & No & 438 & 88.66\% \\\midrule
            
            \multirow{2}{*}{Egocentricity} & Yes & 201 & 40.69\% \\
            & No & 293 & 59.31\% \\
		\bottomrule
	\end{tabular}
\end{table}

\begin{table}
\centering

\caption{Number of respondents assigned to each condition (i.e., who saw each \texttt{privacy statement} and \texttt{auditing statement}, or no \texttt{privacy statement} (bottom row) or no \texttt{auditing statement} (columns 4 and 7).}
\label{tab:scenarios}
\begin{tabular}{l|ccc|ccc}
    \toprule
       & \multicolumn{3}{c}{\textit{For-Profit}} & \multicolumn{3}{c}{\textit{Non-Profit}} \\
       & AG(1) & AG(2) & Ctrl. & AG(1) & AG(2) & Ctrl. \\\midrule
         PG(1) & 22 & 18 & 19 & 19 & 20 & 18 \\
         PG(2) & 17 & 21 & 19 & 18 & 18 & 19 \\
         PG(3) & 18 & 20 & 18 & 19 & 19 & 17 \\
         PG(4) & 21 & 18 & 19 & 20 & 18 & 19 \\
         Ctrl. & -  & -  & 19 & -  & -  & 21 \\
    \bottomrule
\end{tabular}
\end{table}




\subsection{Analysis}
\label{ss:analysis}


We analyzed the open-text questions about \texttt{willingness to donate} and \texttt{trust} (see~\autoref{ss:method-design}) using inductive-thematic open coding. Two researchers independently coded each entry and generated a codebook from a random sample of at least 100 (20.2\%) responses. Then, they composed a final codebook and double-coded all responses. 
Since all responses were double-coded and inconsistencies were resolved, we do not report inter-rater reliability (IRR)~\cite{mcdonald2019reliability}. Ultimately, six responses did not fit this scheme and classified as `Other'. The codebook of the open-text responses is presented in~\autoref{app:code}.




In addition to presenting a descriptive analysis detailing the distribution of responses on the survey items that address our research questions, we construct logistic regression models to analyze factors related to two dependent variables: \texttt{privacy expectations} (RQ1) and \texttt{willingness to donate} (RQ2). For RQ1, the independent variables were the presence of a given privacy and/or auditing statement. The dependent variable was the \texttt{privacy expectation} corresponding to the \texttt{privacy statement} presented to a given respondent. Using this model we compare the privacy expectations of respondents in the experimental conditions (those shown a \texttt{privacy statement}) with those of the control groups. Responses from respondents in an experimental conditions are only modeled in the analysis of the \texttt{privacy statement} they were shown.

For RQ2, the dependent variable was \texttt{willingness to donate}, and the independent variables were the presence of a given \texttt{privacy statement}, the \texttt{privacy expectation} for each guarantee, 
as well as demographics and experiences.
We collapsed the independent variables into a binary measure, designating responses of ``Likely'' or ``Very Likely'' as True and all others as False. We did so to avoid the ambiguity introduced by intermediate scale points and to simplify the statistical model.

We categorized education into two groups: with and without a bachelor's degree. We took ``no bachelor's degree'' and ``less than forty years old'' as the reference categories, respectively. For gender, we took man as the reference category. Technical background and donation history are binary factors, and we took the negative response as the reference category. We built separate regression models for the two recipient entities: for-profit and non-profit.


\subsection{Limitations}
\label{ss:method-limit}

Although the four privacy-preserving guarantees examined in this study (\autoref{s:intro}) cover a broad range of PETs, they are not fully comprehensive and do not capture the full complexity of PETs or real-world threats, which may involve privacy issues beyond the scope of our investigation. 
Furthermore, our approach may not fully reflect the intricacies of real-world donation scenarios, where multiple guarantees may coexist. While such complexities lie outside the primary focus of our research, qualitative investigations have been conducted in prior work~\cite{Aitken2016,mori2016public}, and quantitative exploration of these aspects remains an interesting direction for future research.

Despite our efforts to mitigate misunderstandings—through rigorous cognitive interviews, filtering respondents who reported not fully understanding the scenarios, and controlling for privacy expectations in our statistical analyses—the brief descriptions of each guarantee might still have led to partial or incorrect understanding among respondents. This limitation may have influenced our results.

Additionally, our study relies on self-reported data, which is subject to well-documented biases, known as the privacy paradox, where individuals' expressed privacy concerns often differ from their actual behaviors~\cite{privacyparadox}. To address these limitations, we implemented several methodological safeguards.
Firstly, we employed vignette-based scenarios, which research has shown to effectively mirror real-world behaviors and reduce hypothetical bias~\cite{Redmiles2017ASO}. 
Secondly, our work is informed by prior research from 2019~\cite{8835345} and replicated in 2022~\cite{beyondtheprivacyparadox} that demonstrated crowdsourced samples, specifically from Prolific~\cite{beyondtheprivacyparadox}, well approximate the security and privacy behavior, expectations, and knowledge of the general population for US adults between 18-50 who have at least some college education. 
While there may still be discrepancies between participants' expressed privacy concerns and their actual behaviors, research confirms that this overambitiousness is systematic~\cite{askingforafriend,collectingsurvey}. It establishes survey data as a reliable upper-bound approximation of real-world behavior.

Furthermore, statistically significant results do not inherently eliminate the risk of underlying biases that could skew the findings. Various unmeasured factors not accounted for in our analysis might have influenced the interpretations of the results. Lastly, our respondents were recruited via Prolific, which may limit the generalizability of our findings. The sample might not fully represent the diversity of the U.S. population, nor does it capture the perspectives of individuals from other countries, highlighting a key limitation of this work.

%% file: results.tex
\subsection{RQ1: Understanding of PGs}
\label{ss:result-protection}

We first address RQ1 (see~\autoref{s:intro}): ``How well do people understand and expect privacy guarantees PG(1)--PG(4) and auditing guarantees AG(1)--AG(2)?'' We analyze survey responses and their underlying rationales.

\subsubsection{Analysis Setup}

\begin{figure}[t]
    \centering
    \includegraphics[width=.48\textwidth]{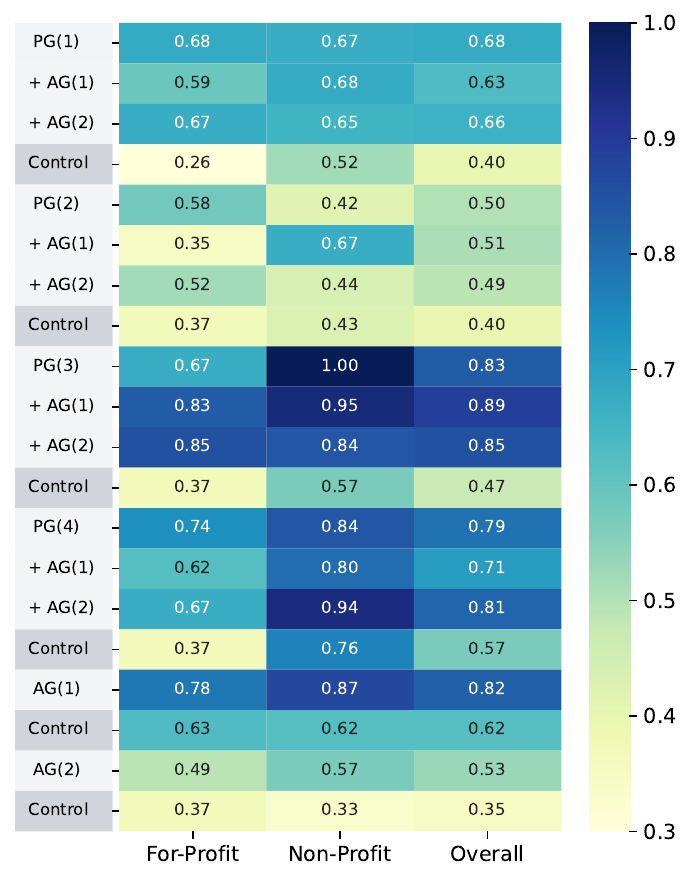}
    \caption{Percentage of respondents who had a positive \texttt{privacy expectation} when shown a particular \texttt{privacy statement} in their donation scenario. The values in the table are the percentage of respondents in a given condition who had the \texttt{privacy expectation} (see~\autoref{fig:sqconcerns}) that corresponded to the \texttt{privacy statement} they were presented. The 3rd column (``overall'') reports results across both entities. For example, the right-most and top-most numerical cell indicates that 68\% of participants in both entity scenarios who were shown PG(1) alone -- with no \texttt{auditing statement} -- expected their data to be anonymized. The agreements are binarized here to present the overall tendency.}
    \label{fig:protection}
\end{figure}

\begin{figure}[ht]
    \centering
    \includegraphics[width=\linewidth]{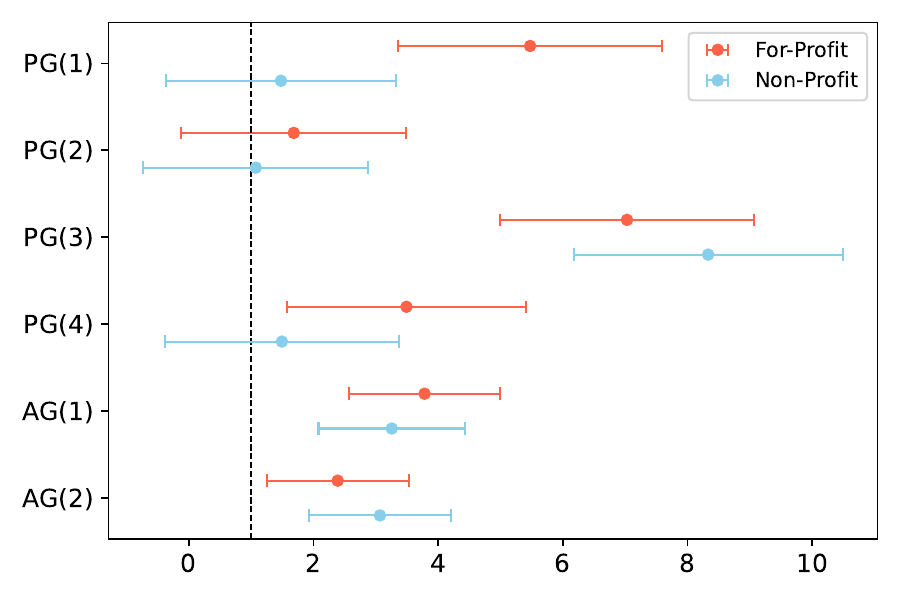}
    \caption{Odd ratios and 95\% confidence intervals of presenting different \texttt{privacy statements} on the respondents' \texttt{privacy expectations} toward the two kinds of recipient entities, for-profit and non-profit.}
    \label{fig:present-stat}
\end{figure}

We compare \texttt{privacy expectations} on PG(1)---PG(4) and AG(1)--AG(2) among groups of respondents who received different privacy statements (listed in \autoref{fig:privacystatements}) and auditing statements (listed in \autoref{fig:auditingstatements}). In~\autoref{fig:protection}, we show the \texttt{privacy expectations} of respondents who were shown different \texttt{privacy statements}.
To assess the statistical significance of the descriptive quantitative results presented above, we used logistic regression to analyze the relationship between the presence of a \texttt{privacy statement} in the scenario and respondents' \texttt{privacy expectations}. We summarize results, which we distinguish for non-profit and for-profit entities, in~\autoref{fig:present-stat}.

\subsubsection{Response Distribution and Statistical Analysis Results}
We find that \textbf{privacy expectations differ based on the data-collecting entity.}
In the control group, which was not shown any \texttt{privacy statement}, 26\%-37\% of respondents expected for-profit entities to provide anonymization, data expiration, access control, purpose restriction, expert auditing, and self-auditing. In contrast, a higher percentage of respondents (43\%-76\%) expected non-profit entities to employ these mechanisms, except for self-auditing (AG2). 
For instance, we note that respondent P430 (who received a for-profit scenario) expected privacy protections even though they were in the control group (and thus received no statement about privacy protection): \emph{``I also assume that the data is looked at in the aggregate and likely no one at the company knows me''}. In the same situation, respondent P53 (who received a non-profit scenario instead) made a stronger assumption: \emph{``I trust that the organization will uphold strict privacy and ethical standards.''}

In the for-profit scenario, all privacy statements, except for the access-control statement (PG2), significantly increase the corresponding privacy expectations ($p$-value < $0.05$).
In the non-profit scenario, where expectations are already high, only the privacy statement about data expiration (PG3) significantly increases expectations (from 57\% to 100\%). Comparing the for-profit and non-profit subsets of the descriptive results (1st and 2nd column of~\autoref{fig:protection}), we observe that presenting a \texttt{privacy statement} was particularly beneficial in for-profit scenarios. Indeed, the presence of a \texttt{privacy statement} raised respondents' \texttt{privacy expectations} to nearly the same level as non-profit scenarios for anonymization (PG1, 68\% and 67\%) and access control (PG2, 58\% and 42\%). It also reduced the gap in expectations for data expiration (PG3, 67\% and 100\%)) and purpose restriction (PG4, 74\% and 84\%)).

Auditing statements significantly increased expectations of corresponding audits in both for-profit and non-profit scenarios ($p$-value < $0.05$).
However, we note that \textbf{incorporating an audit statement does not significantly change \texttt{privacy expectations}}. The statistical significance of the auditing statements in~\autoref{fig:present-stat} indicates that respondents shown the audit statement correctly expected the corresponding type of audit, while respondents that were not shown the auditing statement did not expect this protection. 
As observed in ~\autoref{fig:protection}, presenting an \texttt{auditing statement} does not increase the number of respondents who expect the privacy guarantee implied by the \texttt{privacy statement} they were shown, despite the stronger assurances that audits imply.

\subsubsection{Qualitative Analysis of Privacy Expectations via Open-Answer Responses}

Across for-profit and non-profit scenarios, 69\% and 85\% of respondents, on average, expressed that their positive privacy expectations were formed based on the presence of the privacy statement, belief in privacy obligations of the entity, trust in the entity, or trust in the auditing process.

\PP{Satisfaction with the privacy guarantees.} 16.3\% and 20.8\% of respondents in the for-profit and non-profit scenarios, respectively, were convinced by the received \texttt{privacy statement}. Some respondents, like P56, vaguely stated that the policy looks promising: \emph{``I see their policy, and they have to follow their own policy.''} Others felt explicitly safer and trusted the statement. For example, P494 noted: \emph{``Trust is very important when it comes to medical data. I believe the organization has privacy policies that outline that they collect and will use my data. I also believe the organization will employ security measures to safeguard data.''} Similarly, P555 highlighted the presence of the \texttt{privacy statement}: \emph{``(...) purposely states there is software in place to conserve privacy.''} and P516 felt their information was safe because of it: \emph{``(...) with the privacy protection in place, they are isolating the data they need while basically `throwing out' the rest by putting in under that protection. In essence, my information is safe, and they're only using what they said they'd use.''} P208 also felt safer contributing to research knowing the privacy mechanisms were in place: \emph{``I want to be able to contribute to research to better improve cancer treatment, and I feel safe if my data is protected through the mechanisms above.''}
    
\PP{Belief in legal or reputational obligations.} 16.3\% and 10.6\% of respondents in the for-profit and non-profit scenarios, respectively, believed that entities are forced to protect the donated data due to legislative requirements or reputational concerns. For example, P28 stated regarding for-profits: \emph{``A for-profit organization wouldn't want to violate HIPAA, HITECH laws.''} Additionally, P117 wrote \emph{``I would trust them to do the right thing so they won't face lawsuits.''} Regarding non-profits, P141 noted: \emph{``I trust that the law will restrict any data leaks to third parties.''} 
    
\PP{Trust in the recipient entity.} 13.9\% and 17.5\% of respondents in the for-profit and non-profit scenarios, respectively, expressed a general trust in the recipient entity without specifying reasons. For instance, P70 succinctly said: \emph{``I feel they are reliable and trustworthy''} and P50 stated that \emph{``I assume they take their research seriously, so they would handle the data carefully.''} Other respondents, like P145, reported having no reason not to trust it: \emph{``I have no reason to think they would do anything nefarious with my medical data.''}

\PP{Trust in the auditing process.} 7.8\% and 7.9\% of respondents in the for-profit and non-profit scenarios, respectively, found the auditing statement to add at least some reliability. For example, P202 showed some reservation but found confidence in the public nature of audits: \emph{``I don't fully trust them, but I somewhat do, particularly if audits and verification of results are made public. That said, claiming that the advisory board is external is only partly reassuring, as it's appointed by the institute.''} Additionally, P540 had trust in the entity's data handling because \emph{``they let outsiders audit them''} and P51 (who received the expert-auditing statement) because of \emph{``safelocks and checks in place''}. On the other hand, one interviewee doubted the expert auditing process and stated: \emph{``somebody else says something doesn't mean that it’s real''}. No survey respondent explicitly raised concern towards the auditing process.

Across for-profit and non-profit scenarios, 31\% and 15\% of respondents, on average, expressed negative privacy expectations because they were skeptical of the privacy statement or doubted whether the recipient entity would actually employ PETs as stated. Their qualitative responses offer insights on the underlying reasons:

\PP{Skepticism on the privacy statement and limits of privacy-preserving guarantees.} 21.2\% and 10.5\% of respondents in the for-profit and non-profit scenarios, respectively, expressed general distrust in the feasibility of the privacy statement. In the non-profit scenario, P34 stated that \emph{``no privacy technology is foolproof''}. Similarly, in the for-profit scenario, P45 wrote: \emph{``I don't trust that the privacy-preserving tech would work.''} More precisely, privacy statements were considered too ideal to be fully enforced. Some respondents believed that unauthorized employee access would be unavoidable. For instance, P71 said: \emph{``Why would I trust someone other than my doctor with my medical records? These days especially, I don't trust anyone. There could be a breach or simply people I don't know from a hole in the wall will then have access to all my med records. Insane''!} Some others felt that data breaches and cyberattacks were inevitable. For instance, P83 stated that \emph{``the primary reason for my distrust is due to past news of companies being hacked by people and their data getting leaked''}, while P124 noted that \emph{``corporate data breaches are very common''}, and P80 echoed this sentiment, saying: \emph{``I believe the [intentions] will be good, but data can be hacked''}. We observe that the auditing statement did not substantially instill trust on all accounts --- with the exception of expert auditing for access control in non-profit scenario, and data expiration in for-profit scenario. For example, P474 mentioned: \emph{``They can check their privacy technology all they want but when there is a breach it is done and info is stolen. After it fails then they say sorry and offer monitoring but the info is still stolen.''}

\PP{Doubt on the recipient entity's motivation to employ PETs.} 8.6\% and 3.3\% of participants in the for- and non-profit scenarios, respectively, claimed that these entities are inherently self-serving and lack the motivation to uphold privacy-preserving guarantees or implement such measures at all. Most criticism was directed at for-profit entities. For instance, P55 noted that for-profit entities \emph{``will do what is profitable and not much more than that''}. Similarly, P119 stated \emph{``(...) because it is a for-profit organization. I expect them to cut corners''} and P32 wrote that \emph{``for-profit organization have low standard of morality''}. However, some respondents also expressed concerns about non-profit entities. For example, P190 in the non-profit scenario remarked: \emph{``Medicine has become a business. My data is only useful to them if it helps them make more money. Money comes first before the actual well being of humans.''}

\subsection{RQ2: Willingness to donate health data}
\label{ss:result-donation}

Next, we answer the second research question (RQ2, see~\autoref{s:intro}): ``How does the deployment of privacy guarantees and auditing influence people's willingness to donate their personal health data?'' As in the previous subsection, we statistically analyze the responses and then report the qualitative rationales collected from the open-text questions.

\subsubsection{Analysis Set Up}

\begin{figure}[t]
    \centering
    \includegraphics[width=.48\textwidth]{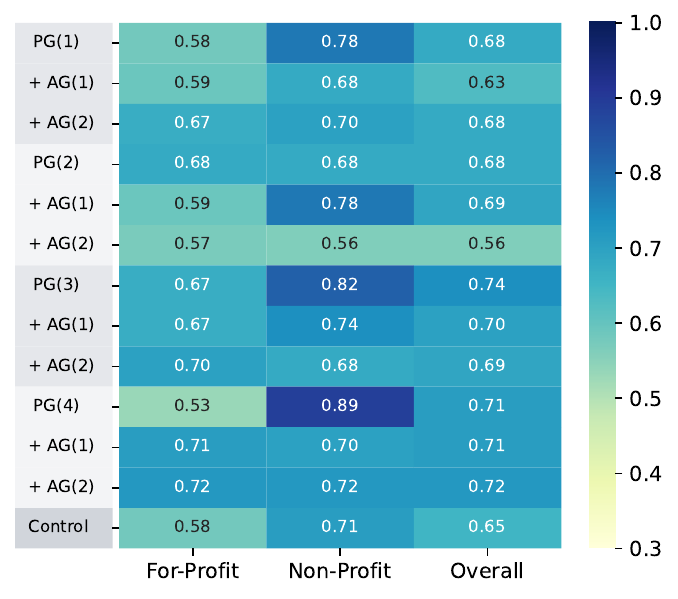}
    \caption{Percentage of respondents willing to donate their personal health data to the recipient entity in each scenario (e.g., the right-most and top-most numerical cell indicates that 68\% of participants in both entity scenarios that with PG(1) presented but no \texttt{auditing statement} were willing to donate.) The agreements are binarized here to present the overall tendency.}
    \label{fig:donation}
\end{figure}

In~\autoref{fig:donation}, we summarize the respondents' \texttt{willingness to donate} their health data to the presented recipient entity. The 1st column (``overall'') reports results for both entity types, the 2nd column reports the results of the subset with for-profit entity, and the 3rd column reports the results of the subset with non-profit entity. The values represent the percentage of respondents willing to donate their heath data to the recipient entity. 

We constructed a logistic regression model to understand the factors that influence \texttt{willingness to donate}. 
We first analyze and confirm the influence of the recipient entity and then separate the participants based on the recipient entities and analyze them in separate models. The results are summarized in~\autoref{tab:donation-stat}. 

In~\autoref{fig:venn-reaction}, we visualize the overlap in experimental group respondent's \texttt{willingness to donate}, \texttt{privacy expectation} for the statement they were shown, and \texttt{trust} that the entity would protect their data as described.

\subsubsection{Response Distribution and Statistical Analysis Results}

\begin{figure}[t]
    \centering
    \includegraphics[width=.48\textwidth]{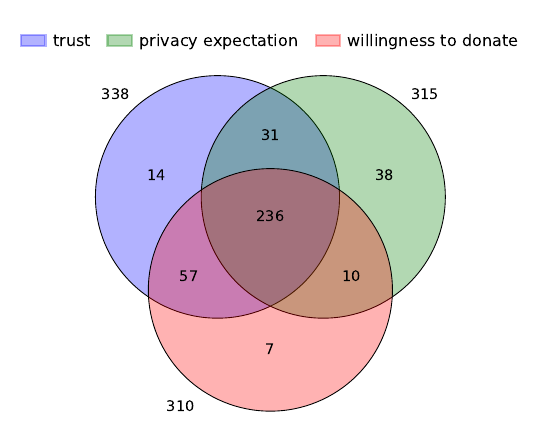}
    \caption{Overlap in respondent's \texttt{willingness to donate}, \texttt{privacy expectation} for the statement they were shown, and \texttt{trust} that the entity would protect their data as described (respondents in the control -- no \texttt{privacy statement} -- are excluded). The numbers outside each circle summarize the total number of respondents who were e.g., willing to donate their data (310). $N=454$.}
    \label{fig:venn-reaction}
\end{figure}

\begin{table}[t]
	\centering
	\caption{Influences of \texttt{privacy expectations} and \texttt{demographics} on \texttt{willingness to donate} of the three participant groups. OR: odd ratio. CI: confidence interval.}
	\label{tab:donation-stat}
	\begin{tabular}{c|lccc}
		\toprule
		Entity & Factor Levels & OR & 95\% CI & $p$-value \\
		\midrule
		Both
		& Entity - FP & 0.38 & [0.24, 0.59] & \textbf{\textless 0.001} \\
		\midrule        %
        %
        %
        %
        %
        %
        %
        %
        %
		\parbox[t]{2mm}{\multirow{19}{*}{\rotatebox[origin=c]{90}{For-Profit}}}
		& \multicolumn{4}{l}{\textit{Privacy Statement}} \\
            & \quad PG(1) & 0.65 & [0.25, 1.70] & 0.381 \\
             & \quad PG(2) & 0.68 & [0.27, 1.75] & 0.429 \\
             & \quad PG(3) & 0.86 & [0.32, 2.28] & 0.756 \\
             & \quad PG(4) & 0.62 & [0.22, 1.70] & 0.352 \\
		& \multicolumn{4}{l}{\textit{Privacy Expectation}} \\
		   & \quad PG(1) & 1.09 & [0.66, 1.80] & 0.741 \\
             & \quad PG(2) & 1.60 & [0.96, 2.69] & 0.073 \\
             & \quad PG(3) & 1.70 & [1.01, 2.88] & \textbf{0.047} \\
             & \quad PG(4) & 3.25 & [1.78, 5.87] & \textbf{\textless 0.001} \\
             & \quad AG(1) & 2.46 & [1.40, 4.27] & \textbf{0.002} \\
             & \quad AG(2) & 1.32 & [0.76, 2.27] & 0.317 \\
		& \multicolumn{4}{l}{\textit{Demographics \& Experiences}} \\
		& \quad Education          & 0.73 & [0.43, 1.26] & 0.264 \\
             & \quad Age                & 1.00 & [0.98, 1.01] & 0.765 \\
             & \quad Gender             & 1.30 & [0.80, 2.12] & 0.304 \\
             & \quad Tech Background    & 1.57 & [0.88, 2.80] & 0.127 \\
             & \quad Egocentricity      & 0.69 & [0.42, 1.14] & 0.144 \\
             & \quad Donation History   & 3.25 & [1.31, 8.01] & \textbf{0.011} \\
		\midrule
        %
        %
        %
        %
        %
        %
        %
        %
        %
		\parbox[t]{2mm}{\multirow{19}{*}{\rotatebox[origin=c]{90}{Non-Profit}}}
		& \multicolumn{4}{l}{\textit{Privacy Statement}} \\
		& \quad PG(1) & 1.27 & [0.52, 3.13] & 0.603 \\
            & \quad PG(2) & 0.94 & [0.37, 2.38] & 0.901 \\
            & \quad PG(3) & 1.17 & [0.45, 3.00] & 0.746 \\
            & \quad PG(4) & 1.54 & [0.61, 3.90] & 0.369 \\
		& \multicolumn{4}{l}{\textit{Privacy Expectation}} \\
		& \quad PG(1) & 1.72 & [1.02, 2.89] & \textbf{0.043} \\
             & \quad PG(2) & 1.46 & [0.87, 2.46] & 0.152 \\
             & \quad PG(3) & 1.27 & [0.73, 2.25] & 0.400 \\
             & \quad PG(4) & 5.21 & [2.61, 10.38] & \textbf{\textless 0.001} \\
             & \quad AG(1) & 2.59 & [1.35, 4.90] & \textbf{0.004} \\
             & \quad AG(2) & 1.17 & [0.68, 2.03] & 0.563 \\
		& \multicolumn{4}{l}{\textit{Demographics \& Experiences}} \\
            & \quad Education         & 0.46 & [0.26, 0.80] & \textbf{0.007} \\
            & \quad Age               & 1.00 & [0.98, 1.02] & 0.897 \\
            & \quad Gender            & 1.16 & [0.69, 1.97] & 0.560 \\
            & \quad Tech Background   & 0.71 & [0.40, 1.26] & 0.237 \\
            & \quad Egocentricity     & 1.36 & [0.83, 2.23] & 0.217 \\
            & \quad Donation History  & 2.25 & [1.06, 4.75] & \textbf{0.033} \\
		\bottomrule
	\end{tabular}
\end{table}

There was no direct relationship between showing a \texttt{privacy statement} and respondents' \texttt{willingness to donate}. However, several \texttt{privacy expectations} positively correlate with \texttt{willingness to donate} for both non-profit and for-profit entities, except in specific cases like access control for both for- and non-profit and data expiration for the non-profit.
Visualized in~\autoref{fig:venn-reaction}, we observe high alignment among the three explored constructs, where 236 (52.0\%) respondents gave all positive responses and 63 (13.9\%) respondents gave all negative responses. Combined with the results of RQ(1), the findings suggest that \textbf{\texttt{willingness to donate} is influenced by \texttt{privacy expectations}, which are, in turn, impacted by \texttt{privacy statements}}. 

Privacy statements and expectations influence donation intentions, but they are not the only factors. While participants were generally less willing to donate to for-profit entities (OR=0.38, $p$-value < $0.001$), other factors are also at play. Looking at control group participants (who received no privacy statements), we found that the difference in \texttt{willingness to donate} between for-profit and non-profit entities was only 8.8\%, despite a much larger 20.3\% difference in \texttt{privacy expectations} between these entities. This discrepancy suggests that \textbf{non-privacy-related factors, such as perceived donation benefits, also influence participants' \texttt{willingness to donate}}.


Furthermore, we observe that  \textbf{respondents with bachelor's degrees are less willing to donate to a non-profit entity (OR = 0.46, $p$-value = $0.007$)}, which may be related to their greater awareness of the complexities and potential vulnerabilities with \texttt{privacy statement}. They have heightened privacy concerns that need to be addressed. For example, P42 wanted to verify that the PET worked visually; P97 thought that only the stored data was encrypted, but the data transaction was not protected the same way; and P236 argued that: \textit{``privacy-preserving technology that works today may not work in a few years''.}

Finally, we found that \textbf{prior donors are more willing to donate to both entities.} Respondents who have donated before were more willing to donate to both entities, with the OR value of 3.25 ($p$-value = $0.011$) and 2.25 ($p$-value = $0.033$) for the for- and non-profit entities, respectively.
These respondents expressed strong willingness of supporting science and recognized the importance of medical data in developing new treatments. For example, P40 claimed: \textit{``I feel that it is important information to share in hopes that they can find better ways to deal with diabetes''.} and P208 responded: \textit{``Breakthroughs in science and moving forward in knowledge, our greatly benefited by such Data Collection.''}

\subsubsection{Qualitative Analysis of Donation Intention Via Open-Answer Responses}

\textbf{Many respondents reasons for donating (or not) included a desire to support research, ego-centric connections, or desire for personal reward.} In the for-profit scenario, 50.2\% of respondents' reasons for donating (or not) fell into these categories, in the non-profit scenario, 53.3\% did.

One non-privacy-related reason for donating included is helping research. For example, P107 wrote that \emph{``there are a number of things I could share to research such as money, time, data, and more. I would be willing to share my medical records to contribute to cancer research in hopes that people will be healthier and to enhance research future medicines''.} P97 stated: \emph{``even if they are for-profit (which I don't like), any closer we get to developing better treatments and/or a potential cure to cancer is something I'd be willing to assist''.} 

Ego-centric reasons, particularly familial connections, also motivated donations. P154 mentioned that the motivation was: \emph{``my dad died when his heart failed on him, and I'd have rather that not had happened''}, and P233 noted that they \emph{``would be willing to help to prevent someone else's mom from dying of a stroke like [theirs] did''.}

However, their existing (negative) experiences with for-profit healthcare entities also deter some of them from making donations. For example, P136 claimed they have heard multiple instances where entities promise not to sell information but circumvent this by using different terminology, effectively still selling the data. P244 also mentioned that their information has been leaked by a clinician in the past. 

Some respondents desired personal reward for donating and thus were not interested in donating in our scenario: e.g., P92 explained that \emph{``I would give them my data if I got paid for it. I would think twice if they want me to donate the data and make no money from it while the researchers will make money off the of research they compiled.''}
An alternative explanation could be the perceived imbalance in cost and profit between the donor and the recipient entity, where the donor provides data for free while the recipient entity profits from its use. For instance, participants P495, P463, and P459 all share this perception. However, we note that some individual participants alternatively were motivated to donate due to egocentricity, even to for-profit entities. For example, P45 \textit{``just lost my father to heart failure thirteen days ago. (...) I would be more than willing to donate my medical records if it helps develop medicines to make hearts function better.''}

\vspace{5px}
\textbf{Respondents expressed negative sentiments toward for-profit entities collecting data.} The desire for personal reward related closely to lack of donation intent for for-profit organizations. In the for-profit scenario, 11.4\% explicitly refused due to negative perceptions of such organizations. For instance, P191 and P202 respectively stated: \emph{``I have a negative connotation with for profit organization''} and \emph{``I feel like this type of organization already profits significantly off a large number of people (occasionally off of me as well) and as such, I do not want to give them direct permission to profit further off of me.''} Furthermore, respondents felt that for-profits already benefit enough and were unwilling to contribute further. For instance, P36 wrote: \emph{``For profit organization makes profit using the data. I'm never going to donate it. They can BUY it from me for a reasonable amount.''} 

\vspace{5px}
Alternatively, a small group of 1.3\% respondents explicitly mentioned that \textbf{they would donate to non-profits but remained concerned about potential data misuse}. For instance, P122 mentions: \emph{`it's always going to be in the back of my head that there's a possibility that it's being sold or taken or used in some way I'm not okay with. Scares me a bit.''} Then, P282 notes: \emph{`it is difficult to police everyone in a non-profit organization. I have seen leaders in organizations act unethically at times, which makes me think that even a well-intended organization cannot fully control the actions of every employee or volunteer they have.''}

\vspace{5px}
\textbf{Privacy-related donation considerations focused on data sensitivity and leakage.} Concerns for not donating were mostly about data leakage, with respondents expressing that they 12.9\% and 13.9\% participants in the for-profit and non-profit scenario, respectively, were protective of their data and hence would not donate at all, as P545 noted that \emph{``there are so many data leaks from so called safe places. No one can anticipate what hackers can do in the future.''} Others cited previous data leak experiences as to why they would not donate, as P290 said: \emph{``my medical data has been breached in the past by a clinician and I had to file a formal grievance against the health care system that employed her. For this reason, because my medical history was abused, I no longer have any trust and will never voluntarily consent to my medical history/information being shared with others.''} Some respondents also expressed that they would only possibly donate to specific entities that have already proven trustworthiness to them, as P357 explained that \emph{``regardless of privacy policy, the probability of a leak or misuse is high. unless it is an organization I have had personal interaction with, or am very familiar with, it is unlikely I would donate my medical records.''} 

Some participants expressed concern about the level of detail in medical data. P155 highlighted this concern: \emph{``I would not feel comfortable with so much of my personal health information being shared with an organization that is not involved with my direct medical care, regardless of whether there is an advisory board or not''}; as did P229: \emph{``It's a lot of detailed information that I'm worried if it somehow gets in the wrong hands, it could reveal a lot of private information about me.''} Contrastingly, some respondents felt comfortable sharing data due to a lack of sensitive information (e.g., \emph{``I don't have mayor illnesses or nothing to hide so Im ok with that''}) or because they already share information with other entities (\emph{e.g., ``Considering I already disclose this information to other organizations (for example, data collection on phone, data compiled through search history), other agencies likely have the information on hand already and so another organization having it is no different''}).

%% file: discuss.tex
Health data donation decisions involve multiple complex factors beyond privacy considerations. Non-privacy factors, such as recipient reputation and perceived societal benefit, often influence willingness to donate. Moreover, the effectiveness of PETs depends on people's belief on these protections. PETs' impact on donation willingness diminishes substantially when individuals misunderstand guarantees or doubt implementation integrity. 

Prior work has highlighted the ways in which statements on explaining PETs can fail due to being vague or inaccurate. Our work extends this understanding by showing that even when people clearly understand the protection guarantees, the perceived effectiveness of these guarantees is often limited by non-privacy-related factors. 
Our participants reported strong pre-existing privacy expectations, \textit{even in the absence of explicit privacy statements}. We observe that these expectations vary strongly based on the profit model of the entity. For for-profit entities, where baseline privacy expectations are lower, strong privacy statements can effectively elevate these expectations among users. In contrast, the impact on non-profit organizations is minimal because users already hold high privacy expectations for these entities.
This disparity in privacy expectations between non-profit and for-profit entities reflects deeper psychological mechanisms that influence how people form judgments about organizational trustworthiness and privacy practices.

\PP{The Halo Effect}
Participants' qualitative responses suggest that their strongly positive privacy expectations for non-profit entities arise from basic moral judgments: non-profits are perceived as more ethical and dedicated to serving the public good because they are not pursuing profit. This findings are consistent with a phenomenon termed ``the halo effect'', which describes how an overall positive impression of a particular entity (e.g., the good work that a non-profit does) can lead to ungrounded assumptions about how the organization operates~\cite{nisbett1977halo,de2024moral}. While prior work in the business literature focuses on how the halo effect influences expectations about business practices such as donation management, our work illustrates that the halo effect also influences people's expectations about the way non-profits handle data privacy. 

However, these positive expectations do not always align with reality. Privacy safeguards are not consistently enforced or widely practiced in non-profit entities; for example, many health organizations rely on broad consent frameworks rather than adopting stricter purpose-restricted consent measures~\cite{kaye2015dynamic}. 
This gap between expectations and reality can result in privacy expectation violations, a major source of privacy concerns as explained by the theory of contextual integrity~\cite{nissenbaum2009privacy}. Indeed, a well-known non-profit mental health helpline shared it's data with a commercial entity, leading to significant public outcry~\cite{levine2022suicide}. Future work may seek to explore how to design privacy communications that set appropriate expectations in the presence of such effects, as a prior work has sought to do in other domains (e.g., \cite{burton2015broken}).

\PP{Privacy Washing}
Conversely, our participants had far lower initial expectations of the privacy practices of for-profit entities due to their perceived singular focus on profit (a ``horn'' effect). As a result, information about data protection guarantees significantly raised their privacy expectations, and in turn, their willingness to donate their data. 

This means for-profit entities can gain competitive advantage~\cite{schaub2015design} by transparently implementing and communicating PETs to counter their trust deficit. Yet this dynamic creates vulnerability to ``privacy washing''~\cite{mainwp2023privacywashing,cirucci2024oversharing}, where organizations make vague or exaggerated privacy claims to deceptively raise expectations. 
A recent incident involving GoodRx exemplifies this problem: despite promising never to share personal health information, it shared sensitive data with third-party advertisers~\cite{fdcgoodrx}.
Such deception ultimately erodes trust and discourages future donations when discovered, deepening the very horn effect that motivated privacy washing.

\PP{Opt in Versus Opt out}
Switching to an ``opt-out'' model for health data donation might seem appealing to overcome donation hesitancy, particularly for for-profit entities struggling against lower baseline trust. However, such approaches face legal barriers in the U.S. HIPAA~\cite{HIPAA} sets a high bar for health data sharing, requiring explicit ``opt-in'' authorization for most research uses.

\PP{The Role of Auditing}
In theory, auditing could serve as a powerful mechanism to counterbalance both the halo effect and privacy washing by providing objective verification of privacy guarantees, regardless of organizational structure. Auditing should transform abstract guarantees into verifiable technical reality. Yet our findings reveal that auditing has surprisingly limited impact on elevating privacy expectations among participants, with the exception of scenarios involving purpose use restrictions and for-profit entities.

This limited effectiveness arises from participants' fundamental skepticism about technological infallibility., and shows a disconnect between expert and public perspectives on verification mechanisms, consistent with findings in prior work~\cite{dechand2019encryption,Oates2018,wu2018tree}. 
While security and privacy experts typically regard auditing as foundational to establishing trust and demonstrating compliance, our participants takes it as merely another layer of protection that could ultimately fail. As one respondent succinctly stated, ``They can guarantee privacy all they want; things still get hacked.''
This skepticism aligns with emerging research suggesting that many users have developed a resigned attitude toward privacy, doubting that true privacy protection is achievable in practice~\cite{Lerner2022}.

\PP{Recommendations for Recipient Entities}
Given participants' skepticism about privacy protections, recipient entities seeking health data donations should consider several approaches to address these concerns. First, they should recognize that users' prior experiences with data leakage and exposure to media coverage of breaches create a sense of inevitability about privacy risks. This fatalistic perspective requires more than just technical solutions.

For technical implementations, recipient entities should consider advanced technologies like multi-party computation (MPC) that can minimize breach impacts by distributing sensitive information across multiple entities, ensuring no single point of failure. However, these systems must be engineered carefully, as any failure could damage trust more severely than if simpler methods were compromised.

When communicating privacy guarantees, recipient entities should move beyond explaining basic techniques like E2EE. Our findings show that participants struggle to understand more complex concepts like auditing, despite their effectiveness. Recipient entities should develop communication approaches that better explain these sophisticated cryptographic concepts in accessible ways that demonstrate their benefits.

For-profit entities in particular should recognize their trust disadvantage and focus on transparent implementation and clear communication of robust privacy protections. At the same time, they should avoid ``privacy washing'' through vague or exaggerated claims, which can erode trust if discovered.

\PP{Recommendations for End-Users}
Our research highlights several key points that individuals should consider before donating their health data. First, donors should evaluate recipient entities based on the data protections they explicitly state in policy documents rather than their organizational structure. The halo effect observed in our study demonstrates how implicit assumptions about certain types of entities' business practices can lead to unfounded assumptions about privacy practices, particularly for non-profit entities.

Second, individuals should avoid assuming privacy protections exist beyond those explicitly stated by the entity requesting data. While entities are legally bound to follow both their explicit promises and applicable privacy laws, the baseline legal protections may be less comprehensive than what users expect for sensitive health data~\cite{pearman2022user}.

Finally, individuals should consider engaging privacy advocates before donating sensitive health data. Such consultation can help identify potential privacy risks that might not be immediately obvious and provide a more balanced perspective on the true privacy implications of donation.

\PP{Future Work}
While our study provides valuable insights into privacy expectations and willingness to donate health data, several promising directions remain for future research. First, researchers could expand beyond our four privacy guarantees and two auditing mechanisms to incorporate a more comprehensive range of PETs, with a particular focus on understanding how advanced and complex guarantees can be effectively communicated to non-technical audiences and how they shape privacy expectations and willingness to donate. This could include exploring the interplay between different privacy guarantees and how they are perceived by individuals.

Second, future work should examine these privacy expectations across more diverse demographic groups, including individuals with a strong distrust of data-handling practices. Identifying and understanding these individuals is critical in fostering confidence in PETs and practices.

Third, our study focused on formal medical data donation scenarios, yet health data is increasingly collected through passive means via consumer devices and casual in-app permissions. As prior work emphasizes the importance of modality, timing, and context in privacy information design~\cite{schaub2015design}, future research could explore how such factors intersect with biases such as the halo effect to influence privacy expectations, trust, and willingness to share sensitive information.

Finally, the increasingly common partnerships between for-profit and non-profit entities in the medical domain warrant dedicated study. Building on prior work~\cite{seberger2021post} that found deep skepticism toward hybrid entities in COVID-19 contact tracing, future work may seek to further interrogate how such partnerships, which are common in the medical domain, further complicate our understanding of people's privacy expectations and their sensitivity to statements of data protection.

%% file: ack.tex
All authors were supported in full or in part by the Max Planck Society during the course of this research. The fourth author additionally acknowledges support from a Google Research Scholar Award ``Aligning Technical Data Privacy Protections with User Concerns'' and NSF Award \# 2429838.

%% file: appendix.tex
\subsection{Survey}
\label{app:survey}

\twovers{Omitted due to space; see the \href{https://arxiv.org/abs/2407.03451}{extended version}~\cite{wang2024role}.}{
In this section, we report the full text of our questionnaire. 
We first ask for a participant themselve or their close relatives have been diagnosed with a disease.
In the rest of the questionnaire, we will choose at random one disease the participant has selected, if any, or one disease the participant has not selected. 
This is labeled in Italic as \textit{disease}. The \textit{entity}, also in Italic, may be either non-profit entity or for-profit entity, chosen at random.

\paragraph{Egocentricity}

\begin{enumerate}
	\item Have \textit{you} ever been diagnosed with any of these diseases?
	\begin{itemize}
		\item Cancer (Breast, Colorectal, Lung, and Prostate)
		\item Diabetes
		\item Heart Failure
		\item Hypertension (High blood pressure)
		\item Stroke
		\item None of the above
		\end{itemize}
	\item 
	Have \textit{your partner and/or close relatives (parents, children, siblings)} ever been diagnosed with any of these diseases?
	\begin{itemize}
		\item Cancer (Breast, Colorectal, Lung, and Prostate)
		\item Diabetes
		\item Heart Failure
		\item Hypertension (High blood pressure)
		\item Stroke
		\item None of the above
		\end{itemize}
\end{enumerate}

\paragraph{Survey Introduction}
Next, we present you with a scenario and ask you how you would respond. Please do your best to imagine yourself in this scenario. There is no right or wrong answer to these questions: we are interested to hear how you would feel about being in this scenario and what you would do.

Imagine that an entity  \textit{entity} wants to develop a new treatment for  \textit{disease}. They need medical data from people with and without  \textit{disease} to develop the treatment. They ask you to donate your medical record to help develop the treatment.

Your medical record contains your: (i) personal information, which may include information about your age, weight, gender, race; (ii) medical history, which may include information about allergies, illnesses, surgeries, immunizations, and results of physical exams and tests; and (iii) medical behavior, which may include information about medicines taken and health habits, such as smoking habits, diet and exercise.

\paragraph{Questions}

We will now ask you a series of questions about this scenario. For your reference, the scenario text will appear at the top of each page.

\begin{enumerate}
	\item How would you rate your understanding of the above scenario?
	\begin{itemize}
		\item Fully understand 
		\item Mostly understand
		\item Partially understand
		\item Not understand
	\end{itemize}
	\item  In this scenario how likely would you be to donate your medical record?
		\begin{itemize}
		\item Very likely 
		\item Likely
		\item Unlikely
		\item Very unlikely
	\end{itemize}
	\item Why your are willing (or unwilling) to share your medical record with this \textit{entity}?
	\item Please rate your agreement with the following statement:
	
	I trust the \textit{entity} will handle my data as described.
		\begin{itemize}
		\item Strongly agree
		\item Somewhat agree
		\item Somewhat disagree
		\item Strongly disagree
		\end{itemize}
		
	\item Please explain why you do (or do not) trust that the \textit{entity} will handle your data as described.
	
	\item In this scenario, how likely do you think it is that each of the following will occur?
	\begin{itemize}
		\item My full name or other personal identifiable information will be linked to the donated medical record.
		\begin{itemize}
	\item Very likely 
	\item Likely
	\item Unlikely
	\item Very unlikely
\end{itemize}
\item The donated medical record will be deleted at a set point in time.
		\begin{itemize}
	\item Very likely 
	\item Likely
	\item Unlikely
	\item Very unlikely
\end{itemize}
\item Any employee at the recipient entity will be able to access the donated medical records. 
		\begin{itemize}
	\item Very likely 
	\item Likely
	\item Unlikely
	\item Very unlikely
\end{itemize}
\item The donated medical records will be used for another purpose without my consent. 
		\begin{itemize}
	\item Very likely 
	\item Likely
	\item Unlikely
	\item Very unlikely
\end{itemize}
\item A group of independent experts will verify whether the privacy-preserving technology works and publish a report on their findings. 
		\begin{itemize}
	\item Very likely 
	\item Likely
	\item Unlikely
	\item Very unlikely
\end{itemize}
\item I will be able to hire someone to verify that my medical record is protected as described.
		\begin{itemize}
	\item Very likely 
	\item Likely
	\item Unlikely
	\item Very unlikely
\end{itemize}
\item If I donate my data, I will meet Albert Einstein.
		\begin{itemize}
	\item Very likely 
	\item Likely
	\item Unlikely
	\item Very unlikely
\end{itemize}
\end{itemize}
\item History In real life, have you ever donated your medical record?
		\begin{itemize}
	\item Yes, I have donated my medical record for  \textit{disease}
	\item Yes, I have donated my medical record for research of other illnesses.  
	\item Yes, I have donated my medical record for non-medical purposes.  
	\item No, I was asked to donate my medical record but I declined to participate.  
	\item No, I have never been asked to donate my medical record.  
	\item I prefer not to answer.  
\end{itemize}
\end{enumerate}

\paragraph{Internet Skills}

\begin{enumerate}
	\item How familiar are you with the following terminology? Please choose a number between 1 and 5 where 1 represents “no understanding” and 5 represents “full understanding” of the term.
	\begin{itemize}
		\item Advanced search 
		\item PDF
		\item Spyware
		\item Wiki
		\item Cache
		\item Phishing
		\end{itemize}
	\item 
	Which of the following best describes your educational background or job field?
	\begin{itemize}
		\item	I have an education in, or work in, the field of computer science, computer engineering or IT
		\item I DO NOT have an education in, nor do I work in, the field of computer science, computer engineering or IT.  
		\item Prefer not to answer.
		\end{itemize}
			\item 
		Have you ever heard any of the following techniques? (select all that apply)
			\begin{itemize}
			\item Differential Privacy
			\item End-to-end encryption
			\item Public-key cryptography
			\item Trusted execution environment (TEE)
			\item Pseudo-party computation 
			\item None of the above
		\end{itemize}
\end{enumerate}

\paragraph{Demographics}

\begin{enumerate}
	\item What year were you born in? (4 digits)
	\item To which gender do you most identify? (Select all that may apply)
			\begin{itemize}
		\item Woman
		\item Man
		\item Non-binary
		\item Prefer not to answer
		\item Prefer to self-describe
	\end{itemize}
			\item What is the highest level of education you have completed?
					\begin{itemize}
			\item Less than high school degree  (1) 
			\item High school graduate (high school diploma or equivalent including GED)  (2) 
			\item	Some college but no degree  (3) 
			\item	Associate's degree  (4) 
			\item	Bachelor's degree  (5) 
			\item	Advanced degree (e.g., Master's, doctorate)  (6) 
			\item	Prefer not to answer  (7) 
						\end{itemize}
\end{enumerate}
}
\subsection{Code Book}
\label{app:code}
\twovers{Omitted due to space; see the \href{https://arxiv.org/abs/2407.03451}{extended version}~\cite{wang2024role}.}{
\PP{Data Donation}
\begin{enumerate}
    \item Irrelevant response
    \item Opt to donate regardless
    \item Opt not to donate regardless
    \item Require compensation
    \item Only donate to a non-profit entity
    \item No donation to a for-profit entity
    \item Need more information
    \item The provided protection is persuasive
    \item Generally, the donation scenario looks good
    \item Not satisfied with the provided protection
    \item Other
\end{enumerate}

\PP{Trustworthy}
\begin{enumerate}
    \item Trust without a specific reason
    \item Irrelevant response
    \item Perfect protection does not exist
    \item The recipient entity has no reason not to protect the data
    \item The recipient entity has no reason to protect the data
    \item A non-profit entity is trustworthy
    \item A for-profit entity is not trustworthy
    \item Need more information
    \item The provided protection is persuasive
    \item Generally, the donation scenario looks good
    \item Not satisfied with the provided statement
    \item Other
\end{enumerate}
}
\subsection{Major Changes Following the Pilot Study}
\label{app:pilot}

We conducted an initial pilot study and supplemented this with a series of cognitive interviews to identify areas for improvement in our survey.  The key modifications included:

\twovers{}{
\begin{figure*}[ht]
        \centering
		\begin{tabular}{|p{.95\textwidth}|}
			\hline
			\textbf{PG(1): Anonymization}\\
			\begin{tabular}{@{}p{.95\textwidth}@{}}
				They want to take care to protect the contributors' identities: only the relevant information from the contributors' records can be requested or stored, while any personal identifiable information will not be requested or stored.
			\end{tabular} \\
			\hline\hline
			\textbf{PG(2): Access control}\\
			\begin{tabular}{@{}p{0.95\textwidth}@{}}
				They want to take care to protect the contributors' data by adopting an access policy: only authorized scientists affiliated with the recipient can access the donated medical records.
			\end{tabular} \\
			\hline\hline
			\textbf{PG(3): Data expiration}\\
			\begin{tabular}{@{}p{0.95\textwidth}@{}}
				They want to take care to protect the contributors' data by setting a data expiration time: all contributors' data will be removed after the expiration time, which is stated when you upload your medical record.
			\end{tabular} \\
			\hline\hline
			\textbf{PG(4): Purpose restriction}\\
			\begin{tabular}{@{}p{0.95\textwidth}@{}}
				They want to take care to protect the contributors' data by restricting the usage of the donated data: the contributors' records can only be processed in the context of requests relevant to this project.
			\end{tabular} \\
			\hline\hline			
			\textbf{AG: Auditing}\\
			\begin{tabular}{@{}p{0.95\textwidth}@{}}
				An independent party will audit the recipient and guarantee that their protection methods will be strictly enforced.
			\end{tabular} \\
			\hline
	\end{tabular}
	\caption{Original PG and AG statements.}
	\label{fig:original_statements}
\end{figure*}
}

\PP{Simplify Statements and Understanding}
The initial version of the PG and AG statements \twovers{can be found in our \href{https://arxiv.org/abs/2407.03451}{full appendix}.}{are listed in~\autoref{fig:original_statements}.} However, we receive feedback from the pilot study that the original statements are too abstract and technical. For example, one participant described the anonymization statement as: ``at first glance it sounds good but I would say it's a little bit big'' and ``I'm kind of stumped about that one but um is it so so big'' for the data expiration statement.

We observed participants struggled to understand the statements. We thus include the understanding question in the survey to filter out participants who do not understand the statements.

\PP{Separate Auditing Statements}
In the original study design, we included one auditing statement focused on external auditing. However, our pilot study revealed that participants had concerns about collusion between external auditors and recipient entities. To address this, we introduced a distinct self-auditing statement reflecting current best practices in data governance frameworks and provides higher level of assurance.

\PP{Presentation Enhancement}
The original survey presented all statements in the same visual format, which our pilot study revealed could cause participants to overlook the privacy statements. To address this issue, we applied a colored background to key statements, creating clear visual distinction from the surrounding text (see~\autoref{fig:sample}).

\PP{Demographics Collection}
Based on pilot study feedback, we refined our demographics collection by removing income-related questions that were present in the original questionnaire. Additionally, we introduced questions about participants' educational background and professional field to better assess their familiarity with computer science concepts and provide more context for interpreting their responses.

\subsection{Justification on Included PETs}
\label{app:justification}
\textbf{Anonymization} is a foundational privacy guarantee aimed at stripping or obfuscating personal identifiers in data so individuals cannot be readily identified. In privacy-preserving systems, anonymization enables beneficial use of sensitive data without exposing identities, though it involves a trade-off between privacy and utility~\cite{x1}. Simple de-identification has proven inadequate: numerous studies show ostensibly anonymized datasets can be re-linked with external information to re-identify individuals~\cite{x2, x3}. To address these risks, stronger techniques such as differential privacy have been developed, providing formal guarantees by injecting statistical noise into query results~\cite{x4}. Anonymization complements other privacy safeguards by reducing data identifiability, ensuring that even if data is accessed or leaked, it carries less risk of exposing personal information.

\textbf{Access Control} restricts data access to authorized users, embodying the principle of least privilege in privacy-preserving systems. In healthcare, this ensures only appropriate medical personnel can view sensitive records. In cloud storage, access control becomes more complex as users must trust external servers; without safeguards, administrators could potentially access confidential data~\cite{x5}. To mitigate this risk, researchers have developed cryptographic approaches like end-to-end encryption that prevent even cloud providers from reading stored data without permission~\cite{x6}. Access control complements other privacy measures by acting as the first line of defense, limiting who can access data and reducing the attack surface for privacy violations.

\textbf{Data Expiration} (also called data retirement or assured deletion) is a privacy guarantee that limits how long sensitive information remains accessible in a system. The idea is that personal data should automatically become irretrievable after it is no longer needed, preventing indefinite retention that could later lead to misuse or breaches. This is particularly important in healthcare, where regulations and ethics mandate retaining patient information only as long as necessary for treatment or research; enforcing expiration helps uphold those limits even on persistent cloud backups. Technically, implementing assured deletion is challenging, but systems like Vanish~\cite{x7} demonstrated a feasible approach: encrypting data with a key that automatically disappears from a global distributed hash table after a set time, thereby making the ciphertext permanently unreadable once the timer expires. The ability to make data “self-destruct” in this way reflects the notion that the right and ability to destroy data are essential to protect fundamental societal goals like privacy and liberty~\cite{x7}. By shrinking the window of exposure, data expiration complements other privacy-preserving measures: even if sensitive data is stolen or improperly accessed, it will not persist indefinitely, greatly limiting long-term privacy risks.

\textbf{Purpose Restriction} limits personal data use to only the specific purposes for which it was collected. In healthcare, this prevents patient data collected for treatment from being repurposed for marketing without consent. Implementation typically involves binding data to metadata about permitted uses and enforcing appropriate checks. Researchers have proposed making purpose a first-class parameter in database systems - as seen in Hippocratic databases that enforce "limited use" principles~\cite{x9}. Purpose restrictions are often supported by accountability mechanisms that verify data was only accessed in ways consistent with its intended purpose~\cite{x10}. This guarantee complements anonymization and access control by ensuring that even authorized data access remains confined to legitimate contexts, thereby maintaining patient trust and meeting legal privacy obligations~\cite{de2022covault}.

\textbf{Auditing} is widely recognized as a cornerstone of privacy-preserving systems, providing accountability and fostering user trust in sensitive data environments. In domains like healthcare, where patient records are highly sensitive, robust audit trails help maintain compliance with privacy regulations (e.g., HIPAA) and deter misuse of data. For example, secure logging in electronic health record systems can record an “irrefutable trace” of each user's activity, discouraging unauthorized access and aiding breach investigations~\cite{xx1}. Likewise, in cloud storage, auditing mechanisms (often involving independent or cryptographically verifiable checks) ensure that service providers uphold data confidentiality and integrity commitments. Third-party or automated audits allow customers to “assess and expose risk,” ultimately giving providers incentives to improve their services and reducing the risk of data lapses over time~\cite{xx2}. Major academic venues have highlighted these needs, with numerous systems incorporating secure audit logs and accountability frameworks as core features. For instance, researchers have proposed using cryptographic attestations as tamper-evident audit records so that clients (or regulators) can verify a cloud provider's behavior~\cite{xx3}, and patient-centric accountability schemes that track how medical data is shared to expose any inappropriate access. By enabling such transparent oversight, auditing complements privacy-preserving techniques -- ensuring that even as data remains protected (via encryption or access control), any access or policy violation leaves a verifiable trail. This capability is essential for legal compliance and for maintaining public confidence in both healthcare information systems and cloud data services.